# Effects of polarization modulation induced by electro-optic modulators in fiber-based setups

Frédéric Du-Burck, Karim Manamanni, Tatiana Steshchenko, Amine Chaouche Ramdane and Vincent Roncin

*Abstract*— Using the Jones formalism, it is shown that electro-optic modulators used for phase modulation generate a modulation of the output polarization induced by the difference of phase modulation depth along the crystallographic axes of the modulator. We study two consequences of this polarization modulation in the fiber setups, limiting the performance of high sensitivity measurement devices. The first one is its partial conversion into a residual amplitude modulation (RAM) within any component presenting polarization dependent losses (PDL). The second one is a new effect that consists of the distortion of the signal detected at the output of a fiber cavity. The theoretical expressions of the detected signals are computed in each case.

## I. Introduction

A PERSISTENT issue with electro-optic modulators (EOMs) used to achieve beam phase modulation is their propensity to generate a residual amplitude modulation (RAM) simultaneously with the phase modulation. In spectroscopy or laser frequency stabilization, the demodulation of this RAM leads to a non-zero baseline superimposed to the detected signal [1]. The sensitivity of the RAM generation process to environmental parameters, especially temperature, results in a slow evolution of this offset. RAM also degrades the signal-to-noise ratio of the detected signals by transferring the low-frequency technical noise of the laser at the detection frequency [2]. Finally, the RAM generates a distortion of the detected signal [3]. Hence, RAM is a technical issue that significantly affects the performance of accurate measurement systems.

Several methods have been reported in the literature to compensate RAM or its consequences [4-14]. In particular, a common solution that was initially demonstrated in [15] is to realize a control loop driving the birefringence of the crystal by a DC voltage superimposed on the RF modulation signal.

In waveguide-type fiber coupled electro-optic modulators, the origin of RAM is essentially related to the crystal birefringence [16]. The analysis of this effect is based on the calculation of the modulator output signal for an input laser polarization misaligned with respect to the birefringent crystal axes [15]. As we will show hereafter, this model does not predict the direct generation of an amplitude modulation at the EOM's output, but a modulation of the beam polarization. It is the analysis of the latter by polarization-sensitive components that results in the RAM appearance. This is particularly the case for fibered setups in which the components generally have polarization-dependent losses (PDL). Thus, the generation in EOMs of a polarization modulation leads to the reappearance of RAM within each fiber component presenting PDL.

Furthermore, independently of RAM generation, we have experimentally shown in [17] that the polarization modulation may lead to the distortion of the signal detected at the output of a fiber cavity.

In this letter, using Jones formalism, we analyze in Section II the generation of a residual polarization modulation (RPM) in EOM introduced by a misalignment of the input beam polarization with its principal axes. Section III shows the partial conversion of this RPM into RAM by using components presenting PDL. Section IV discusses the rejection of RAM with a servo loop that superimposes a DC voltage on the modulator. Finally, Section V analyzes the distortion of the output signal of a fiber cavity for an input beam with RPM.

## II. Generation of residual polarization modulation

We consider the MPX-LN-0.1 phase modulator from iXblue Photonics [18]. It is a "X-cut" type transverse configuration $LiNbO_3$ fibered EOM, in which the x-axis is the ordinary axis and the z-axis is the extraordinary one. The light propagates along the y-axis and the electric field is applied along the z-axis (Fig. 1a). The ordinary (along the x-axis) and extraordinary (along the z-axis) indices are given by [15]

$$n_x = n_o - \frac{1}{2} n_o^3 r_o E_z, \qquad (1)$$

$$n_z = n_e - \frac{1}{2} n_e^3 r_e E_z, \qquad (2)$$

where $n_o$, $n_e$ are the ordinary and extraordinary indexes of the undisturbed material respectively, and $r_o$, $r_e$ are the corresponding electro-optical coefficients. The electric field is obtained by applying an electric voltage $V$ between both faces of the crystal.

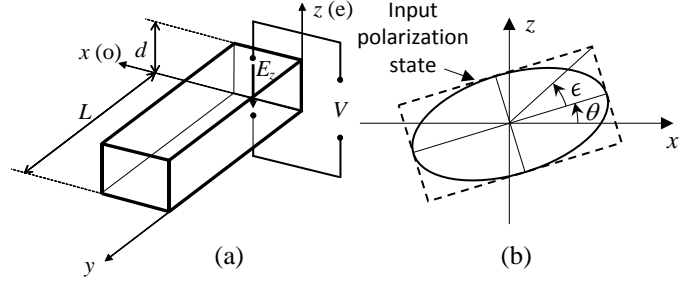

Fig. 1. The EOM considered for the modeling. (a) Structure of the EOM; (b) The input polarization state.

Ideally, beam polarization at the modulator input should be linear and aligned with one of its principal axes (the z-axis for this modulator). Input and output fibers are polarization-maintaining (PM) fibers whose axes coincide with those of the modulator. In practice, residual misalignments, splicings, connectors and PM fiber features lead to a slightly elliptical and misaligned input polarization state with respect to modulator principal axes (Fig. 1b). The polarization extinction ratio (PER) is approximately 25 dB. In the proper base of the ellipse, the optical field at the modulator input is characterized by the Jones vector

$$q_0 = \begin{pmatrix} \cos \epsilon \\ i \sin \epsilon \end{pmatrix} E_0 \, e^{i \, 2\pi \nu_0 t} \tag{3}$$

where $E_0$ is the field amplitude and $\nu_0$ the optical frequency. In the modulator base $(0x, 0z)$, the Jones vector takes the form

$$Q_0 = \begin{pmatrix} \cos \chi \\ \sin \chi \, e^{i \phi} \end{pmatrix} E_0 \, e^{i \, (2\pi \nu_0 t + \zeta)} \tag{4}$$

with

$$\cos \chi = \sqrt{\cos^2 \theta \, \cos^2 \epsilon + \sin^2 \theta \, \sin^2 \epsilon}, \tag{5}$$

$$\sin \chi = \sqrt{\sin^2 \theta \, \cos^2 \epsilon + \cos^2 \theta \, \sin^2 \epsilon}, \tag{6}$$

$$\phi = \arg \left\{ \frac{\sin \theta \, \cos \epsilon + i \, \cos \theta \, \sin \epsilon}{\cos \theta \, \cos \epsilon - i \, \sin \theta \, \sin \epsilon} \right\}, \tag{7}$$

$$\zeta = \arg\{\cos \theta \, \cos \epsilon - i \, \sin \theta \, \sin \epsilon\}. \tag{8}$$

where $\theta$ is the angle between the major axis of the input elliptical polarization and the x-axis. The angle $\phi$ is related to the ellipticity of the polarization state.

The action of the modulator on the field is described by the matrix

$$\mathbb{M} = \begin{pmatrix} e^{i \, \Phi_x} & 0 \\ 0 & e^{i \, \Phi_z} \end{pmatrix} \tag{9}$$

where $\Phi_x$ and $\Phi_z$ represent the phase shifts imposed by the modulator along x-axis and z-axis respectively. For a voltage applied to the modulator

$$V(t) = V_{dc} + V_{rf} \, \sin(2\pi F_{mod} t), \tag{10}$$

these phase shifts at the modulation frequency $F_{mod}$ are

$$\Phi_x = A_x + B_x \sin(2\pi F_{mod} t), \tag{11}$$
$$\Phi_z = A_z + B_z \sin(2\pi F_{mod} t), \tag{12}$$

with the quantities

$$A_x = -\frac{2\pi}{\lambda} L \, n_o + \frac{\pi}{\lambda} \frac{L}{d} n_o^3 \, r_o \, V_{dc}, \tag{13}$$

$$B_x = \frac{\pi}{\lambda} \frac{L}{d} n_o^3 \, r_o \, V_{rf}, \tag{14}$$

$$A_z = -\frac{2\pi}{\lambda} L \, n_e + \frac{\pi}{\lambda} \frac{L}{d} n_e^3 \, r_e \, V_{dc}, \tag{15}$$

$$B_z = \frac{\pi}{\lambda} \frac{L}{d} n_e^3 \, r_e \, V_{rf}. \tag{16}$$

The field at the modulator output is thus given by the Jones vector

$$Q_M = \begin{pmatrix} \cos \chi \\ \sin \chi \, e^{i\,[(\phi+A_z-A_x)+(B_z-B_x)\sin(2\pi F_{mod} t)]} \end{pmatrix} E_0 \, e^{i\,(2\pi \nu_0 t+\zeta)} \, e^{i\,[A_x+B_x \sin(2\pi F_{mod} t)]} \qquad (17)$$

where the ellipticity of the polarization state is now determined by the time dependent term $(\phi + A_z - A_x) + (B_z - B_x)\sin(2\pi F_{mod} t)$. This means that the EOM generates a RPM resulting from the different values of the modulation indices along x-axis and z-axis. More precisely, (17) shows that its magnitude is determined by the difference of both indices $B_z - B_x$. This RPM disappears in the ideal case of a linear input polarization ($\epsilon = 0$) which is perfectly aligned to a principal axis of the modulator ($\theta = 0$ or $\theta = 90°$). In this case, $\chi = 0$ or $\chi = 90°$ and the polarization of the field at the modulator output remains linear and purely phase modulated with modulation index $B_x$ ($\chi = 0$) or $B_z$ ($\chi = 90°$).

Fig. 2 shows the result of the time evolution of the polarization modulation at the output of the EOM iXblue Photonics MPX-LN-0.1 during one period. These polarization states are computed for a linear input polarization misaligned by 3.2° from the z-axis of the EOM (*i.e.* PER = 25 dB).

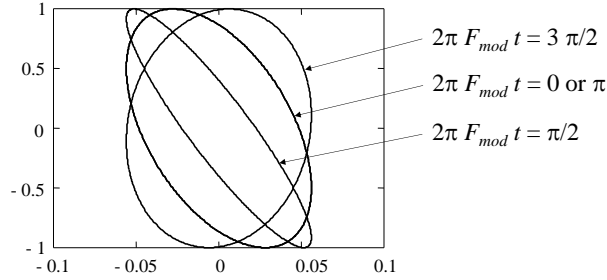

Fig. 2. Time evolution of the polarization state at the EOM output during one period of the modulating signal for a linear input polarization misaligned by $\theta = 3.2°$ with respect to z-axis. According to the characteristics of the EOM, the modulation indices along principal axes are $B_x = 0.28$ and $B_z = 0.91$.

### III. RAM GENERATION

Standard fiber components present polarization dependent losses (PDL) with typical value of about 0.1 dB. Therefore, they analyze the RPM generated by the EOM and produce RAM.

Let us consider for instance that the field is attenuated by a factor $a$ along a direction making an angle $\xi$ with the x-axis (PDL = $1/a^2$). The Fourier component $F_{mod}$ of the photocurrent generated by a photodiode monitoring this field is

$$Y = -2\,|E_0|^2 \cos\chi \sin\chi \cos\xi \sin\xi \,(1-a^2)\, J_1(B_z - B_x) \sin(\phi + A_z - A_x) \sin\psi \qquad (18)$$

where $\psi$ is the detection phase. As a result, for the general case, a DC level is superimposed to the useful detected signal. Note that without PDL ($a = 1$), there is no offset ($Y = 0$) because there is no RAM, although the modulator produces RPM. Thus, RAM is not intrinsic to the modulator and depends on the characteristics of the fiber components ($\xi, a$) within the setup.

Fig. 3 gives the detected RAM level measured at the output of iXblue EOM behind a polarization beam splitter (PBS) as a function of the modulation index $B_z$. We have also plotted in Fig. 3 the theoretical evolution according to (18) using the characteristics of the modulator. The fit of the amplitude is obtained with a linear input polarization misaligned by 1.6° from the z-axis of the EOM (PER = 31 dB).

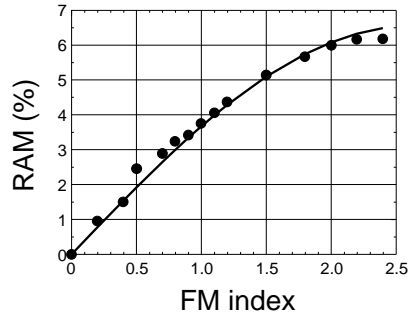

Fig. 3. RAM level normalized by the beam optical power as a function of modulation index. Points are the experimental data and the continuous curve is the theoretical evolution from (18) for $\theta = 1.6°$ and $\epsilon = 0$.

## IV. RAM COMPENSATION

Expression (18) suggests a method of RAM compensation [15]. From (13) and (15), the sine function argument in (18) is equal to

$$\phi - \frac{2\pi}{\lambda} L(n_e - n_o) + \frac{\pi}{\lambda} \frac{L}{d} (n_e^3 r_e - n_o^3 r_o) V_{dc} \tag{19}$$

which depends on the DC voltage $V_{dc}$ superimposed on the RF modulation signal. The RAM compensation scheme is shown in Fig. 4. A PBS is placed at the EOM output and a coupler is connected at the PBS output. The demodulated RAM from one output port of the coupler gives the error signal of a servo loop that controls the voltage $V_{dc}$ in order to maintain the error signal at 0. It should be noted that the PBS is an essential component of this setup because the servo loop rejects RAM but not RPM. Consequently, without PBS, RAM reappears at the output of any component introducing PDL. In the setup of Fig. 4, an ideal PBS produces a perfect linear output polarization and converts RPM to RAM. Then, the servo loop cancels definitely the RAM in the setup.

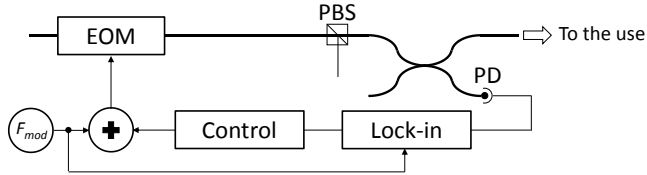

Fig. 4. Basic arrangement for RAM compensation in fiber setup.

## V. DISTORTION OF A FIBER CAVITY OUTPUT SIGNAL

In this section, we examine the effect of the RPM generated by the EOM on the signal detected at the output of a single-mode fiber ring-cavity realized with a fiber coupler and a fiber loop of length $L$. Such cavities are used for instance for laser frequency stabilization [17] or in fiber-optic gyroscopes [19]. In these applications, any distortion of the detected signal at the cavity output introduces a frequency shift limiting the accuracy of measurements.

In order to show clearly the origin of the effect on the detected signal, we consider the simple case for which the cavity has two linear and orthogonal polarization eigenstates due to the linear birefringence of the fiber that constitutes the ring [20]. It is also assumed that the cavity input beam does not present any RAM, *i.e.* there is no component with PDL between the modulator and the cavity input.

The birefringence of the cavity results in a slow axis $Os$ and a fast axis $Of$ associated to indices $n_s$ and $n_f$ respectively. Those axes define the cavity base with the corresponding propagation constants $\beta_s(\nu) = 2\pi \nu n_s/c$ and $\beta_f(\nu) = 2\pi \nu n_f/c$. The Jones matrix of the cavity in this base is

$$\mathbb{C} = \begin{pmatrix} S(\nu) & 0 \\ 0 & F(\nu) \end{pmatrix} \tag{20}$$

in which cavity transfer functions along both principal axes are introduced:

$$S(\nu) = \frac{\sqrt{(1-\gamma)\kappa} - (1-\gamma)\sqrt{\eta}\, e^{-i\beta_s(\nu) L_c}}{1 - \sqrt{(1-\gamma)\kappa\eta}\, e^{-i\beta_s(\nu) L_c}}, \tag{21}$$

$$F(\nu) = \frac{\sqrt{(1-\gamma)\kappa} - (1-\gamma)\sqrt{\eta}\, e^{-i\beta_f(\nu) L_c}}{1 - \sqrt{(1-\gamma)\kappa\eta}\, e^{-i\beta_f(\nu) L_c}}. \tag{22}$$

$\gamma$ and $\kappa$ are respectively the insertion loss and the fraction of input intensity transmitted by the same guide of the coupler and $\eta$ represents the field intensity decrease over one turn in the ring. Coefficients $\gamma$, $\kappa$, and $\eta$ are assumed to be independent from polarization.

The field at the modulator output can be developed according to its Fourier components in the form

$$Q_M = \sum_{n=-\infty}^{+\infty} Q_M^{(n)} \tag{23}$$

where

$$Q_M^{(n)} = \begin{pmatrix} \cos\chi\ e^{i A_x} J_n(B_x) \\ \sin\chi\ e^{i(\phi + A_z)} J_n(B_z) \end{pmatrix} E_0\ e^{i[2\pi(\nu_0 + n F_{mod})t + \zeta]} \tag{24}$$

is the Jones vector associated to the Fourier component $\nu_0 + n F_{mod}$ written in the modulator base $(0x, 0z)$. This field is injected into the cavity. Each Fourier component $Q_M^{(n)}$ of the input field is expressed in the cavity base and transformed by the matrix (20) in order to obtain the output field of the cavity. The resulting Jones vector describing the output polarization expressed in the cavity base is then

$$Q_C = \sum_{n=-\infty}^{+\infty} \begin{pmatrix} S_n \times \sigma_n \\ F_n \times \varphi_n \end{pmatrix} E_0\ e^{i[2\pi(\nu_0 + n F_{mod})t + \zeta]} \tag{25}$$

where $F_n = F(\nu_0 + n F_{mod})$ and $S_n = S(\nu_0 + n F_{mod})$. Quantities $\sigma_n$ and $\varphi_n$ are given by

$$\sigma_n = \cos\varrho \cos\chi \, e^{i\,A_x} J_n(B_x) + \sin\varrho \sin\chi \, e^{i\,(\phi+A_z)} J_n(B_z), \qquad (26)$$
$$\varphi_n = -\sin\varrho \cos\chi \, e^{i\,A_x} J_n(B_x) + \cos\varrho \sin\chi \, e^{i\,(\phi+A_z)} J_n(B_z), \qquad (27)$$

where $\varrho$ is the angle between the slow axis of the cavity ($Os$) and the x-axis of the modulator. We deduce the signal detected at the output of the cavity with the phase detection $\psi$:

$$Y(\nu_0) = \frac{|E_0|^2}{2} \left\{ \sum_{n=-\infty}^{+\infty} \left( S_n S_{n+1}^* \sigma_n \sigma_{n+1}^* e^{i\psi} + S_n S_{n-1}^* \sigma_n \sigma_{n-1}^* e^{-i\psi} \right) \right. \\ \left. + \sum_{n=-\infty}^{+\infty} \left( F_n F_{n+1}^* \varphi_n \varphi_{n+1}^* e^{i\psi} + F_n F_{n-1}^* \varphi_n \varphi_{n-1}^* e^{-i\psi} \right) \right\}. \qquad (28)$$

The detected signals calculated from (28) in different situations are drawn in Fig. 5. The free spectral range of our cavity is 1 MHz and its finesse is 43.5 (mode width of 23 kHz). The modulation frequency is $F_{mod} = 2$ kHz. The refractive index difference between both polarization axes is $10^{-6}$. The dashed curve corresponds to the case where the EOM input beam has a linear polarization aligned on its z-axis and when x-axis and z-axis coincide with principal axes of the cavity. In that configuration the EOM does not generate any RPM. As the cavity input polarization is aligned on one of its principal axes, only one family of modes is observed, without distortion.

As compared to the previous situation, the signal with open circles is obtained by rotating the principal axes of the cavity by 30° with respect to those of the EOM. The second family of cavity modes appears. Since the EOM does not produce RPM, no distortion is observed.

The solid line curve is obtained from the previous situation, but with a linear polarization at the EOM input beam rotated by 20° with respect to its principal axes. Since the EOM output beam now carries RPM, a significant distortion of the detected signal is observed.

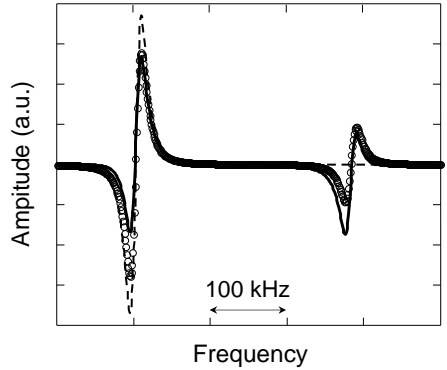

Fig. 5. Signals detected at the output of a fiber cavity for different polarization states at the EOM input (see text). Dashed curve: $\theta = 90°$, $\varrho = 90°$; Open circles: $\theta = 90°$, $\varrho = 60°$; Solid line curve: $\theta = 70°$, $\varrho = 60°$.

We previously observed experimentally this effect for a cavity with a mode linewidth of 800 kHz [17]. We estimated that the zero-crossing shift of the detected signal at the cavity output limited the frequency stability in the $10^{-12}$ range. Our current cavity has much narrower modes (23 kHz). From (28), we find that for an alignment of the polarization axes at the EOM and cavity inputs to within 3°, and a detection phase optimized to maximize the detected signal amplitude to within 3° (distortion depends on detection phase) leads to a zero-crossing offset of ±15 Hz, *i.e.* limits the stability at $1.5\,10^{-13}$.

Note that the introduction of the RAM compensation device described in Section III cancels the RPM and thus eliminates the distortion of the signal detected at the cavity output.

## VI. Conclusion

We have shown in this letter that the difference between the phase modulation indices along the principal axes of an EOM leads to the generation of an RPM associated with the phase modulation. This can limit the accuracy of high sensitivity fiber-based measurement systems in two ways. First, by the conversion from RPM to RAM due to PDL in fibered components. Second, by the distortion of the signals detected at the output of fiber cavities. To our knowledge, this latter effect that we have also experimentally observed [17], has never been described in literature. A RAM compensation device can suppress the RPM and thus the distortion.